\documentclass[journal,comsoc]{IEEEtran}
\usepackage[utf8]{inputenc}

\usepackage{fourier}
\usepackage{tikz}
\usepackage{amsmath,amssymb}
\usepackage{bm}
\usepackage{pgfplots}
\usepackage{cases}

\usepackage{verbatim}

\usepackage{tikz}
\usepackage{algorithm}
\usepackage{algpseudocode}

\pgfplotsset{compat=1.14}

\def \Y {\mathbf Y}

\def \W {\mathbf W}

\def \I {\mathbf I}

\def \w {\mathbf w}

\def \s {\mathbf s}

\def \x {\mathbf x}
\def \y {\mathbf y}
\def \z {\mathbf z}
\def \c {\mathbf c}

\def \w {\mathbf w}

\def \h {\mathbf h}
\def \a {\mathbf a}
\def \b {\mathbf b}

\def \bxi {\boldsymbol \xi}
\def \Ka {K_{\mathrm{a}}}
\def \dof {\mathrm{D}}

\def \BBCH {B_{\mathrm{BCH}}}
\def \BID {B_{\mathrm{ID}}}
\def \Btot {B_{\mathrm{tot}}}

\newtheorem{definition}{Definition}

\usepackage{calrsfs}
\DeclareMathAlphabet{\pazocal}{OMS}{zplm}{m}{n}
\DeclareMathOperator*{\argmin}{\arg\!\min}
\DeclareMathOperator*{\argmax}{\arg\!\max}

\title{Tensor-Based Modulation \\for Unsourced Massive Random Access}
\author{Alexis Decurninge, Ingmar Land and Maxime Guillaud%
\thanks{The authors are with the Mathematical and Algorithmic Sciences Laboratory, Paris Research Center, Huawei Technologies France. e-mail:~\texttt{\emph{firstname.lastname}@huawei.com}.}%
}

\begin{document}
\maketitle

\begin{abstract}
We introduce a modulation for unsourced massive random access whereby the transmitted symbols are  \mbox{rank-1} tensors constructed from Grassmannian sub-constel\-lations.
The use of a low-rank tensor structure, together with tensor decomposition in order to separate the users at the receiver, allows a convenient uncoupling between multi-user separation and single-user demapping.
The proposed signaling scheme is designed for the block fading channel and multiple-antenna settings, and is shown to perform well in comparison to state-of-the-art unsourced approaches.
\end{abstract}

\section{Introduction}

Massive random access, whereby a large number of transmitters communicate with a single receiver, constitutes a key design challenge for future generations of wireless systems.
The considered scenarios typically consider sporadic traffic with small payloads; furthermore, only a fraction of the transmitters are active at a given time.
In that context, it is desirable to let users transmit without any prior resource request (grant-free).
At the physical layer, this requires a departure from the design assumptions prevailing in current cellular systems \cite{wu19}.

The grant-free random access problem has recently been revisited taking massive connectivity into account (see \cite{liu18}).
One classical approach consists in a functional split at the receiver between activity detection and channel estimation on one hand (typically, based on user-specific pilot sequences) and multi-user equalization and demapping of the information-bearing symbols on the other hand.
Another approach lies in ALOHA schemes and their extensions, such as slotted coded ALOHA \cite{Paolini15}.

Recently, the emergence of unsourced random access \cite{polyanskyi17} has sparked a renewed interest in the problem.
In this paradigm, the identity of the active transmitters is not associated with a specific waveform at the physical layer.
Theoretical analysis of the unsourced scenario has been done in \cite{polyanskyi17} in the single-in-single-output (SISO) case for an additive white Gaussian noise (AWGN) channel and extended to  quasi-static Rayleigh fading in \cite{kowshik19}.
Several practical schemes have been proposed for this scenario.
For the SISO AWGN case, \cite{fengler19a} proposed a scheme close to sparse regression codes proposed in \cite{barron_joseph11} where the idea is to see the unsourced access as a very large compressed sensing problem where messages are encoded through sparse vectors.
In order to enable reasonable decoding complexity, the linear compressed sensing mapping is split into blocks while the messages are encoded by a binary outer code.
In \cite{pradhan19}, the authors combined the compressed sensing approach with a multi-user coding scheme, also allowing a low-complexity decoder.
Relaxing the AWGN hypothesis and assuming Rayleigh fading, \cite{kowshik19} proposes a scheme based on a low-density parity check (LDPC) code using a belief propagation decoder.
The MIMO case with Rayleigh fading is addressed in \cite{fengler19} as well as in \cite{shyianov20} wherein encoders inspired from compressed sensing were used while adapting the decoder to the MIMO setup.


In this article, we propose a modulation suitable for massive multiple access, where multi-user multiplexing is handled through the use of tensor algrebra. 
Specifically, each user transmits a sequence that is associated to a rank-one tensor.
The receiver observes the linear combination of these signals weighted by the respective channel realizations, which itself can be interpreted as a tensor summing a number of rank-one components equal to the number of active users.
This structure allows the receiver to separate the users using a classical tensor decomposition, without requiring a separate activity detection or channel estimation step; the channels are estimated jointly with the data by the receiver.
The benefits of the proposed approach are:
\begin{itemize}
    \item user separation can be performed by the receiver without relying on pilot sequences, thus circumventing the difficult problem of pilot sequence design for grant-free access, and without involving any knowledge about the discrete nature of the modulation;
    \item the proposed scheme applies to a broad class of multiple-access channels (including AWGN and block-fading) and generalizes to multiple-antenna receivers while benefiting from spatial diversity, without relying on any assumption on the fading distribution.
\end{itemize}

\section{Unsourced Massive Random Access Background and Channel Model}

We consider the transmission from $K$ single-antenna transmitters to a single receiver equipped with $N$ antennas.
Let us consider a block of $T$ channel uses, during which we assume that only $\Ka\ll K$ users are active (where $\Ka$ is still typically large) and simultaneously transmit a payload of $B$ information bits each.
The set of active users is a random subset of the $K$ users (hence its cardinality $\Ka$ is a random variable) and, following the unsourced random access paradigm \cite{polyanskyi17}, we will work under the assumption that all users use the same constellation $\mathcal{C}=\{\c_1,\dots,\c_{2^B}\}$ containing $2^B$ elements.
Under this assumption, the receiver can only decode the messages up to a permutation over the user indices\footnote{Note that the identity of the transmitting user can be included in the message (e.g. in the form of $\lceil \log_2 K \rceil$ bits). In that case, the total number of users $K$ has an impact on the achieved spectral efficiency since the data payload is reduced to $B - \lceil \log_2 K \rceil$ per user. Here, we ignore that aspect and focus on the unsourced problem.}.

Let $\h_k\in \mathbb{C}^{N}$ denote the channel from user $k$.
We assume in this paper a block-fading model whereby the channel realizations remain constant over the considered block of $T$ channel uses, and is a priori unknown to both the transmitters and the receiver.
Let us further assume without loss of generality (w.l.o.g.) that the active users are indexed by $1,\ldots,\Ka$.
We let $\s_k\in\mathcal{C}\subset \mathbb{C}^T$ denote the sequence of complex baseband symbols transmitted by user $k$ over $T$ channel uses, and $\W\in\mathbb{C}^{T\times N}$ the noise realization.
The users are assumed block-synchronous, therefore the signal $\Y\in\mathbb{C}^{T\times N}$ received by the $N$ antennas over the $T$ channel accesses can be written as $\Y = \sum_{k=1}^{\Ka} \s_k\h_k^T + \W$.
Let $\y, \w \in \mathbb{C}^{TN}$ denote the respective vectorized versions of $\Y$ and $\W$. We can rewrite the received signal using the Kronecker product operator (denoted by $\otimes$) as
\begin{equation}
\label{eq:receive_kron}
\y = \sum_{k=1}^{\Ka} \s_k\otimes \h_k + \w.
\end{equation}

\section{Tensor-based Modulation (TBM)}
\label{sec:tbm}
\subsection{Tensor Structure}
In this work, we propose to design the constellation $\mathcal{C}$ according to a tensor construction.
Here, we merely consider tensors to be multi-dimensional data structures, which can be seen as the generalization of matrices to dimensions greater than 2 (see \cite{comon_mag14} for an introduction).
Specifically, let us consider a complex-valued tensor of order $d$ (which can be construed as $d$-dimensional array of complex scalars) of dimensions $T_1,\dots, T_d$.
Note that the $\prod_{i=1}^d T_i$ scalars forming the tensor can also be stored sequentially in a vector (see \cite[Sec.~2.4]{kolda09}).
The corresponding vectorization operator defines an isomorphism between the space of $(T_1,\dots, T_d)$-dimensional tensors and the space of $(\prod_{i=1}^d T_i)$-dimensional vectors, endowed with the respective sum operations.
For notational convenience, throughout the paper we will use the vectorized representation while referring to algebraic arguments applying in the space of tensors.

In the proposed approach, we assume that the blocklength $T$ can be factored as $T = \prod_{i=1}^d T_i$ for some $d\geq 2$ and $T_1, \ldots, T_d \geq 2$, and constrain $\s_k$ to be the vectorized representation of a rank-1 tensor of dimensions $T_1,\dots, T_d$, characterized by the existence of vectors $\x_{i,k}\in\mathbb{C}^{T_i}$ for $1\leq i\leq d$, such that
\begin{equation}
\label{eq:kronecker}
\s_k = \x_{1,k}\otimes \dots \otimes \x_{d,k}\in\mathbb{C}^{\prod_{i=1}^d T_i}=\mathbb{C}^{T}.
\end{equation}
We further constrain each $\x_{i,k}$ to be an element of a sub-constellation $\mathcal{C}_i$ defined as a discrete subset of $\mathbb{C}^{T_i}$, i.e. $\x_{i,k}\in\mathcal{C}_i\subset\mathbb{C}^{T_i}$.
The resulting vector constellation $\mathcal{C}$ is a discrete subset of $\mathbb{C}^{T}$ comprised of all possible combinations of elements of the sub-constellations, i.e.
\begin{equation}
\mathcal{C} = \Big\{\x_1\otimes\dots\otimes\x_d: \ \x_1\in\mathcal{C}_1, \ \dots \ , \ \x_d\in\mathcal{C}_d\Big\}.
\end{equation}
Substituting \eqref{eq:kronecker} in \eqref{eq:receive_kron}, the received signal becomes 
\begin{equation}
\label{eq:receive2}
\y = \underbrace{\sum_{k=1}^{\Ka} \x_{1,k}\otimes \dots\otimes \x_{d,k} \otimes \h_k}_{\triangleq \y_0} + \w \ \in \mathbb{C}^{TN},
\end{equation}
where we let $\y_0$ denote the noise-free received signal.

A tensor is said to be rank-$r$ whenever $r$ is the smallest integer such that the tensor can be written as a sum of $r$ rank-1 tensors \cite[Sec.~3.1]{kolda09}.
Considering eqs.~\eqref{eq:kronecker} and \eqref{eq:receive2}, note that each user transmits a rank-1 tensor of order $d$, while $\y_0$ is the vector representation of a tensor of order $d+1$ and dimensions $T_1,\dots, T_d, N$ having rank at most $\Ka$.

\subsection{Tensor Decomposition, Identifiability and User Separation}
\label{sec:identifiability}

The proposed modulation design is motivated by the fact that the decomposition of a tensor into a sum of rank-1 components (known as the canonical polyadic decomposition, or CPD) is unique up to a permutation over the components under mild conditions.
Furthermore, tensors of order 3 or more can attain high rank even for moderate tensor sizes (conditions linking the maximum rank to the tensor size differ significantly from the matrix case, and will be detailed below).
This hints at the possibility for the TBM to achieve a high degree of multiplexing, while using the CPD to separate the signal components related to each user.

Let us consider in more detail the maximum rank and unique decomposibility conditions of the noise-free tensor in eq.~\eqref{eq:receive2}.
We have the following definition:
\begin{definition}[CPD uniqueness and rank]
\label{def:continuous_identifiability}
The $(T_1,\dots, T_d, N)$-dimensional tensor represented by $\y_0$ admits a unique rank-$\Ka$ CPD if for any set 
$\big\{\x_{1,k}^{\prime}\in\mathbb{C}^{T_1},\dots,\x_{d,k}^{\prime}\in\mathbb{C}^{T_d},\h_k^{\prime}\in\mathbb{C}^N,1\leq k\leq \Ka^{\prime}\big\}$ with $\Ka^{\prime}\leq \Ka$ such that
\begin{equation}
\sum_{k=1}^{\Ka^{\prime}} \x_{1,k}^{\prime}\otimes\dots\otimes \x_{d,k}^{\prime}\otimes \h_k^{\prime} = \y_0,
\end{equation}
then it holds that $\Ka^{\prime}=\Ka$ and there exists a permutation $\sigma$ such that $\x_{1,k}^{\prime}\otimes\dots\otimes \x_{d,k}^{\prime}\otimes \h_k^{\prime} = \x_{1,\sigma(k)}\otimes \dots\otimes \x_{d,\sigma(k)} \otimes \h_{\sigma(k)}$.
\end{definition}

According to \cite{chiantini14}, there exists an upper bound $\overline{R}$ to the rank of uniquely decomposable tensors, in the sense that the set of rank-$\Ka$ tensors for which the CPD is not unique has measure zero if $\Ka < \overline{R}$.
Let us restate the main result of \cite{chiantini14} using our notations.
For tensors of size $T_1,\dots,T_d,N$, let us assume w.l.o.g. that $T_1\geq T_2\geq \dots\geq T_d$ and define 
\begin{equation}
\label{eq:generic_rank}
R^0 = \Big\lceil{\frac{TN}{N+\sum_{i=1}^{d}(T_i-1)}} \Big\rceil
\end{equation}
(which is known as the \emph{expected generic rank}), as well as 
\begin{eqnarray}
R^1 &=& 2 - N + N\prod_{i=2}^dT_i - \sum_{i=2}^d (T_i-1), \quad \text{and} \\
R^2 &=& 1 + T - \sum_{i=1}^d (T_i-1). 
\end{eqnarray}
According to \cite[Th. 1.1]{chiantini14}, we have\footnote{Note that the proof mechanism used in \cite{chiantini14} limits the result to tensors of size $N\prod_i T_i \leq 15000$, however there is no indication that it does not hold more generally. Note also that we have omitted a few exceptions applying to some specific tensor sizes from our restatement of \cite[Th. 1.1]{chiantini14}.}
\begin{numcases}{\overline{R}=}
R^1 -1 & for $T_1\geq  R^1$ \label{eq_unbalanced1} \\
R^2 -1 & for $N\geq R^2$ \label{eq_unbalanced2} \\
R^0 & otherwise, \label{eq_subgeneric}
\end{numcases}
where eqs.~\eqref{eq_unbalanced1} and  \eqref{eq_unbalanced2} correspond to the unbalanced tensor size case of \cite[Th. 1.1]{chiantini14}.
Note that for both unbalanced cases, it holds that $\overline{R}\leq R^\mathrm{0}$, i.e. the balanced case favors higher rank with respect to the unbalanced case.
The practical consequence of the above result is that in the noise-free case, performing a CPD allows the receiver to separate users with high probability whenever $\Ka<\overline{R}$.

Note that Definition~\ref{def:continuous_identifiability} considers tensor decomposition in a continuous domain (i.e. $\x_{i,k}\in\mathbb{C}^{T_i}$).
A definition more directly related to the communication problem at hand, in the sense that it takes into account the discrete nature of the $\mathcal{C}_i$, is as follows:
\begin{definition}[Discrete identifiability]
\label{def:discrete_identifiability}
The noise-free received tensor is \textit{identifiable in the discrete case} if for any set $\big\{\x_{1,k}^{\prime}\otimes\dots\otimes\x_{d,k}^{\prime}\in\mathcal{C},\h_k^{\prime}\in\mathbb{C}^N,1\leq k\leq \Ka^{\prime}\big\}$ with $\Ka^{\prime}\leq \Ka$ and
\begin{equation}
\sum_{k=1}^{\Ka^{\prime}} \x_{1,k}^{\prime}\otimes\dots\otimes \x_{d,k}^{\prime}\otimes \h_k^{\prime} = \y_0
\end{equation}
then it holds that $\Ka^{\prime}=\Ka$ and there exists a permutation $\sigma$ such that $\x_{i,k}^{\prime}=\x_{i,\sigma(k)}$ and $\h_k^{\prime}=\h_{\sigma(k)}$.
\end{definition}

Clearly, Definition~\ref{def:discrete_identifiability} is appropriate for the communications problem at hand\footnote{We remark that because of the unsourced nature of the scheme, the case of several users transmitting the same payload leads to a failure in the discrete identifiability of $\y_0$.
However, as noted in \cite{polyanskyi17}, this occurs with low probability when the payload size $B$ is large, and can be completely avoided if the payload includes a unique user identifier.}, while the CPD uniqueness condition is unnecessarily strong (there might be tensors which are discretely identifiable but do not admit a unique CPD).
Definition~\ref{def:continuous_identifiability} is nonetheless relevant, since i) it constitutes a sufficient condition for the discrete identifiability, and ii) it is simpler than Definition~\ref{def:discrete_identifiability} since it is independent from the design of the discrete constellation $\mathcal{C}$. 

\subsection{Design of the sub-constellations}
\label{sec:constellation_design}
We now discuss the design of the information-bearing sub-constellations $\mathcal{C}_1,\dots,\mathcal{C}_d$. 
Observe that the rank-1 tensor $\x_{1,k}\otimes \dots\otimes \x_{d,k} \otimes \h_k$ is equal to $\alpha_1 \x_{1,k}\otimes \dots\otimes \alpha_d\x_{d,k} \otimes \alpha_{d+1}\h_k$ with $\alpha_1,\dots,\alpha_{d+1}\in\mathbb{C}$ whenever $\prod_{i=1}^{d+1} \alpha_i = 1$.
This indicates that, even if the CPD perfectly recovers the rank-1 component associated to each transmitter, the information-bearing components $\x_{1,k},\dots,\x_{d,k}$ can only be retrieved up to a set of $d$ complex scalar multiplicative coefficients, effectively providing $d$ parallel (non-interfering) non-coherent SISO block-fading channels to each active user.
In order to account for this scalar indeterminacy, each sub-constellation $\mathcal{C}_i$ can either i) embed at least one reference symbols, or ii) rely on a Grassmannian codebook design suitable for the single-user non-coherent block-fading case \cite{ZhengTse2002Grassman,TWC2020_cubesplit}.




\section{Multi-User Receiver}
\subsection{Maximum Likelihood (ML) Decoder}
Let us consider the joint multi-user ML detection problem under the assumption that the noise $\w$ is Gaussian with i.i.d. coefficients.
For the sake of clarity, we will first assume that the number of active users $\Ka$ is known to the receiver. 
The channel realizations $\h_k$ can be considered as nuisance parameters here, and the ML estimator writes
\begin{equation}
\label{eq:ml}
\left\{\hat{\x}_{i,k}\right\} = \argmin_{\substack{\big\{\x_{i,k}\in\mathcal{C}_i\big\}}}\min_{\big\{\h_{k}\in\mathbb{C}^{N}\big\}} \left\|\y - \sum_{k=1}^{\Ka} \x_{1,k}\otimes \dots\otimes \x_{d,k}\otimes \h_k \right\|_2^2.
\end{equation}
The optimization over $\{\h_{k}\}$ is a least squares problem, and therefore has a closed form solution.
Solving the discrete problem in \eqref{eq:ml} via an exhaustive search, however,  requires $2^{B\Ka}$ evaluations of the objective function, which makes it a task of formidable complexity.

\subsection{Two-Step Decoder}
\label{sec:twostep_decoder}
To circumvent the complexity issue, we propose to exploit the tensor structure of the proposed modulation, which allows a simpler two-step decoding process, as follows.
\subsubsection*{Multi-user separation}
First, an approximate CPD with $\Ka$ components is performed, in order to recover an approximate version (denoted by $\hat{\z}_{i,k}$) of the $\x_{i,k}$; specifically, \eqref{eq:ml} is relaxed to yield the rank-$\Ka$ tensor approximation problem
\begin{equation}
\left\{\hat{\z}_{i,k},\hat{\h}_{k}\right\} = \argmin_{\substack{\z_{i,k}\in\mathbb{C}^{T_i},\  1\leq i\leq d \\ \h_{k}\in\mathbb{C}^{N}}} \left\|\y - \sum_{k=1}^{\Ka} \z_{1,k}\otimes \dots\otimes \z_{d,k}\otimes \h_k \right\|_2^2.
\label{eq:cpd}
\end{equation}

\subsubsection*{Single-user demapping}
The second step consists in performing single-user demapping independently for each user, i.e. for each $1\leq k\leq \Ka$ solve the discrete problem
\begin{equation}
\argmin_{\substack{\x_{i,k}\in\mathcal{C}_i, \  1\leq i\leq d\\ \h_k\in\mathbb{C}^N}} \left\|\hat{\z}_{1,k}\otimes\dots\otimes \hat{\z}_{d,k}\otimes \hat{\h}_k - \x_{1,k}\otimes\dots\otimes \x_{d,k}\otimes \h_k\right\|_2^2.
\label{eq:single-user}
\end{equation}
Solving \eqref{eq:single-user} over $\h_k$ (see details in Appendix~\ref{app:single-user}) indicates that the problem is separable into $d$ instances of the minimum chordal distance demapping problem typical of non-coherent modulations:
\begin{equation}
\hat{\x}_{i,k} = \argmax_{\x_{i,k}\in\mathcal{C}_i} \frac{\left|\x_{i,k}^H\hat{\z}_{i,k}\right|}{\|\hat{\z}_{i,k}\| \|\x_{i,k}\|}.
\label{eq:chordal}
\end{equation}
Solving \eqref{eq:chordal} can still be complex if an exhaustive search over the elements of $\mathcal{C}_i$ is performed.
However, this complexity can be significantly decreased through the use of structured constellations such as the one from \cite{TWC2020_cubesplit}.

\subsection{Random User Activation}
In the random access scenario, the number of active users $\Ka$ is random and unknown to the base station. 
This can be addressed by performing the approximate CPD \eqref{eq:cpd} using an upper bound $\overline{\Ka}$ to $\Ka$ (assuming that the user activation probability is known, $\overline{\Ka}$ can be chosen such that $\overline{\Ka} \geq\Ka$ is fulfilled with arbitrarily high probability).
The subsequent demapping of the $\overline{\Ka}$ rank-1 tensors resulting from the approximate CPD will yield $\overline{\Ka}$ messages, among which at most $\Ka$ correspond to actually transmitted messages.
One option is then to discard the messages based on power thresholding on $|\s_k|$; note however that, even if the noise level can be assumed low, this option is not well mathematically justified, because of the lack of a tensor equivalent to the Eckart–Young theorem \cite{Draisma_best_rank_K_approx_tensors_RMS18}.
Another option is to use a binary code to add redundancy at the transmitter, and to check whether the decoded binary sequences actually fulfill the code constraints.
In this case, the choice of the code should emphasize the error detection capability, since it will be used by the receiver to discard messages that do not correspond to an active user, i.e. for which the demapper output is close to uniform i.i.d. bits.

\section{Achievable Degrees of Freedom}
In order to establish the number of degrees of freedom (DoF) achievable by the proposed TBM, let us consider the case of asymptotically high signal-to-noise ratio (SNR), i.e. we assume that the noise-free signal $\y_0$ is available at the receiver.
The DoF achievability proof relies on a random coding argument, in the sense that we assume that the elements of the sub-constellations $\mathcal{C}_i$ are independently drawn from an absolutely continuous distribution.
We will also assume that the channel realizations $\h_k$ are drawn from an absolutely continuous distribution.
These two assumptions ensure that the received tensor is generic, for which case the CPD is almost surely unique whenever $\Ka<\overline{R}$, according to the results cited in Section~\ref{sec:identifiability}.

\subsection{Per-User DoF}
As already pointed out, the uniqueness of the CPD is defined in terms of the rank-1 components, however each factor $\x_{j,k}$ can only be recovered up to a complex scalar coefficient.
Nonetheless, the uniqueness result can be extended to the sub-constellations by considering that the $\x_{j,k}$ are representatives of Grassmannian variables.
More specifically, for all $1\leq i \leq d$, let $G(T_i,1)$ denote the Grassmannian of lines in dimension $T_i$, i.e. the set of 1-dimensional linear subspaces of $\mathbb{C}^{T_i}$, and let us consider $\x_i$ as a representative of the corresponding point in $G(T_i,1)$ (see \cite{ZhengTse2002Grassman} for background information on Grassmannian codebooks).
Note that every rank-1 component $\x_{1,k}\otimes \dots\otimes \x_{d,k} \otimes \h_k$ maps to a unique point of $G(T_1,1) \times \cdots \times G(T_d,1) \times G(N,1)$.
Consequently, whenever $\Ka<\overline{R}$, the proposed scheme allows the noise-free communication of $d$ complex Grassmannian variables in $G(T_1,1) \ldots G(T_d,1)$ respectively, for each active user (the component in $G(N,1)$ corresponds to the channel, and does not carry information).
As shown in \cite{ZhengTse2002Grassman}, a variable in $G(T,1)$ has $T-1$ DoF.
Summing across the sub-constellations indicates that the TBM approach allows the noise-free communication of $\sum_{i=1}^d (T_i-1)$ DoF per active user.

\subsection{Sum-DoF}
The total DoF achieved in the system is obtained by summing up the per-user DoF; therefore, it is a function of $\Ka$. 
Since every active user contributes $\sum_{i=1}^d (T_i-1)$ DoF up to the maximum number of users for which the rank-$\Ka$ tensor is almost surely identifiable, the sum-DoF is
\begin{equation}
\dof_{\textrm{TBM}}(\Ka) =  \Ka\sum_{i=1}^d (T_i-1) \quad \textrm{for} \quad \Ka \leq \overline{R} -1.
\end{equation}
Clearly, the highest sum-DoF is attained for $\Ka = \overline{R} -1$, and is $\dof_{\textrm{TBM}}(\overline{R} -1) =  (\overline{R} -1)\sum_{i=1}^d (T_i-1)$.
Moreover, using the fact that $\Ka \leq \overline{R}-1 \leq R^0-1 < \frac{TN}{N+\sum_{i=1}^{d}(T_i-1)} $, we can upper bound the DoF independently from the tensor size by
\begin{equation}\label{eq:dof_upperbound}
\dof_{\textrm{TBM}}(\Ka)< N(T-\Ka).
\end{equation}

Another upper bound to the sum-DoF can be obtained by allowing the $\Ka$ transmitters to cooperate; in that case, the considered set-up is equivalent to a non-coherent point-to-point MIMO channel with $\Ka$ transmit antennas and $N$ receive antennas, which has an achievable DoF of \cite{ZhengTse2002Grassman}
\begin{equation}\label{eq:dof_ptp}
\dof_{\textrm{coop}}(\Ka) = M^*(T-M^*)
\end{equation}
where $M^*=\min(\Ka,N,\lfloor T/2\rfloor)$.
If we assume that $N<\min(\Ka,T/2)$, the cooperative bound from eq.~\eqref{eq:dof_ptp} yields $\dof_{\textrm{coop}}(\Ka) = N(T-N)$; comparing with eq.~\eqref{eq:dof_upperbound} therefore highlights a loss of at least $N(\Ka-N)$ DoF due to the uncoordinated access.

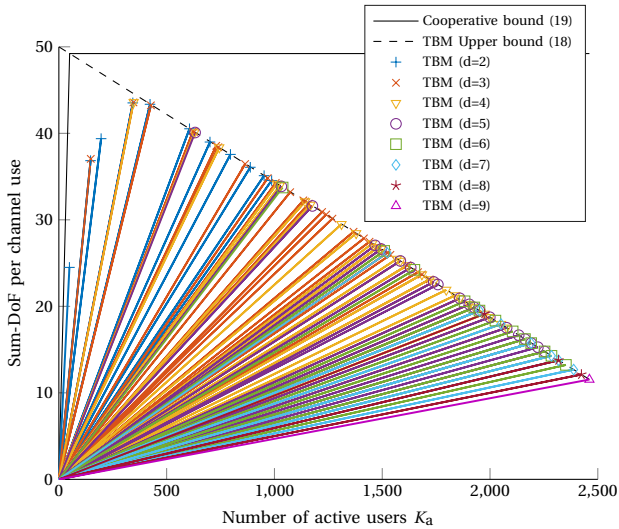
\begin{figure}[ht]
\centering
\scriptsize
%
%
\definecolor{mycolor1}{rgb}{0.00000,0.44700,0.74100}%
\definecolor{mycolor2}{rgb}{0.85000,0.32500,0.09800}%
\definecolor{mycolor3}{rgb}{0.92900,0.69400,0.12500}%
\definecolor{mycolor4}{rgb}{0.49400,0.18400,0.55600}%
\definecolor{mycolor5}{rgb}{0.46600,0.67400,0.18800}%
\definecolor{mycolor6}{rgb}{0.30100,0.74500,0.93300}%
\definecolor{mycolor7}{rgb}{0.63500,0.07800,0.18400}%
\definecolor{mycolor8}{rgb}{0.75000,0.00000,0.75000}%
\begin{tikzpicture}

\begin{axis}[%
width=2.820833in,
height=2.265625in,
at={(1.129375in,0.72875in)},
scale only axis,
xmin=0,
xmax=2500,
xlabel={Number of active users $\Ka$},
ymin=0,
ymax=50,
ylabel={Sum-DoF per channel use},
axis x line*=bottom,
axis y line*=left,
legend style={at={(0.566747,0.594969)},anchor=south west,legend cell align=left,align=left,draw=white!15!black,nodes={scale=0.8, transform shape}}
]
\addplot [color=black,solid]
  table[row sep=crcr]{%
1	0.9996875\\
50	49.21875\\
2461	49.21875\\
};
\addlegendentry{Cooperative bound \eqref{eq:dof_ptp}};

\addplot [color=black,dashed]
  table[row sep=crcr]{%
1	49.984375\\
50	49.21875\\
2461	11.546875\\
};
\addlegendentry{TBM Upper bound \eqref{eq:dof_upperbound}};

\addplot [color=mycolor1,solid,line width=0.75pt,forget plot]
  table[row sep=crcr]{%
1	0.035\\
987	34.545\\
1	0.036875\\
952	35.105\\
1	0.040625\\
888	36.075\\
1	0.0471875\\
796	37.56125\\
1	0.055625\\
701	38.993125\\
1	0.066875\\
606	40.52625\\
1	0.1025\\
423	43.3575\\
1	0.126875\\
343	43.518125\\
1	0.2009375\\
196	39.38375\\
1	0.250625\\
147	36.841875\\
1	0.5\\
49	24.5\\
};
\addplot [color=mycolor1,mark size=2.0pt,only marks,mark=+,mark options={solid}]
  table[row sep=crcr]{%
987	34.545\\
952	35.105\\
888	36.075\\
796	37.56125\\
701	38.993125\\
606	40.52625\\
423	43.3575\\
343	43.518125\\
196	39.38375\\
147	36.841875\\
49	24.5\\
};
\addlegendentry{TBM (d=2)};

\addplot [color=mycolor2,solid,line width=0.75pt,forget plot]
  table[row sep=crcr]{%
1	0.0134375\\
1720	23.1125\\
1	0.0140625\\
1684	23.68125\\
1	0.014375\\
1666	23.94875\\
1	0.0153125\\
1616	24.745\\
1	0.016875\\
1538	25.95375\\
1	0.0171875\\
1523	26.1765625\\
1	0.018125\\
1481	26.843125\\
1	0.0190625\\
1441	27.4690625\\
1	0.0196875\\
1415	27.8578125\\
1	0.0209375\\
1367	28.6215625\\
1	0.02375\\
1269	30.13875\\
1	0.0246875\\
1240	30.6125\\
1	0.0253125\\
1221	30.9065625\\
1	0.0275\\
1159	31.8725\\
1	0.028125\\
1142	32.11875\\
1	0.0284375\\
1134	32.248125\\
1	0.0309375\\
1073	33.1959375\\
1	0.0340625\\
1006	34.266875\\
1	0.0359375\\
969	34.8234375\\
1	0.0421875\\
864	36.45\\
1	0.051875\\
740	38.3875\\
1	0.0528125\\
730	38.553125\\
1	0.0640625\\
627	40.1671875\\
1	0.0646875\\
622	40.235625\\
1	0.10125\\
427	43.23375\\
1	0.1259375\\
346	43.574375\\
1	0.2503125\\
148	37.04625\\
};
\addplot [color=mycolor2,mark size=2.0pt,only marks,mark=x,mark options={solid}]
  table[row sep=crcr]{%
1720	23.1125\\
1684	23.68125\\
1666	23.94875\\
1616	24.745\\
1538	25.95375\\
1523	26.1765625\\
1481	26.843125\\
1441	27.4690625\\
1415	27.8578125\\
1367	28.6215625\\
1269	30.13875\\
1240	30.6125\\
1221	30.9065625\\
1159	31.8725\\
1142	32.11875\\
1134	32.248125\\
1073	33.1959375\\
1006	34.266875\\
969	34.8234375\\
864	36.45\\
740	38.3875\\
730	38.553125\\
627	40.1671875\\
622	40.235625\\
427	43.23375\\
346	43.574375\\
148	37.04625\\
};
\addlegendentry{TBM (d=3)};

\addplot [color=mycolor3,solid,line width=0.75pt,forget plot]
  table[row sep=crcr]{%
1	0.0084375\\
2077	17.5246875\\
1	0.00875\\
2051	17.94625\\
1	0.009375\\
1999	18.740625\\
1	0.0096875\\
1975	19.1328125\\
1	0.0103125\\
1927	19.8721875\\
1	0.010625\\
1904	20.23\\
1	0.01125\\
1860	20.925\\
1	0.0115625\\
1839	21.2634375\\
1	0.0121875\\
1797	21.9009375\\
1	0.013125\\
1739	22.824375\\
1	0.0134375\\
1720	23.1125\\
1	0.0140625\\
1684	23.68125\\
1	0.0153125\\
1616	24.745\\
1	0.0159375\\
1584	25.245\\
1	0.01625\\
1568	25.48\\
1	0.0178125\\
1495	26.6296875\\
1	0.018125\\
1481	26.843125\\
1	0.01875\\
1454	27.2625\\
1	0.020625\\
1379	28.441875\\
1	0.0225\\
1311	29.4975\\
1	0.0271875\\
1167	31.7278125\\
1	0.028125\\
1142	32.11875\\
1	0.033125\\
1025	33.953125\\
1	0.03375\\
1012	34.155\\
1	0.0515625\\
744	38.3625\\
1	0.06375\\
629	40.09875\\
1	0.125625\\
347	43.591875\\
};
\addplot [color=mycolor3,mark size=2.0pt,only marks,mark=triangle,mark options={solid,rotate=180}]
  table[row sep=crcr]{%
2077	17.5246875\\
2051	17.94625\\
1999	18.740625\\
1975	19.1328125\\
1927	19.8721875\\
1904	20.23\\
1860	20.925\\
1839	21.2634375\\
1797	21.9009375\\
1739	22.824375\\
1720	23.1125\\
1684	23.68125\\
1616	24.745\\
1584	25.245\\
1568	25.48\\
1495	26.6296875\\
1481	26.843125\\
1454	27.2625\\
1379	28.441875\\
1311	29.4975\\
1167	31.7278125\\
1142	32.11875\\
1025	33.953125\\
1012	34.155\\
744	38.3625\\
629	40.09875\\
347	43.591875\\
};
\addlegendentry{TBM (d=4)};

\addplot [color=mycolor4,solid,line width=0.75pt,forget plot]
  table[row sep=crcr]{%
1	0.0065625\\
2253	14.7853125\\
1	0.006875\\
2222	15.27625\\
1	0.0071875\\
2191	15.7478125\\
1	0.0075\\
2162	16.215\\
1	0.0078125\\
2133	16.6640625\\
1	0.0084375\\
2077	17.5246875\\
1	0.009375\\
1999	18.740625\\
1	0.01\\
1951	19.51\\
1	0.0103125\\
1927	19.8721875\\
1	0.010625\\
1904	20.23\\
1	0.01125\\
1860	20.925\\
1	0.0128125\\
1758	22.524375\\
1	0.013125\\
1739	22.824375\\
1	0.015\\
1632	24.48\\
1	0.0159375\\
1584	25.245\\
1	0.0178125\\
1495	26.6296875\\
1	0.0184375\\
1467	27.0478125\\
1	0.026875\\
1176	31.605\\
1	0.0328125\\
1032	33.8625\\
1	0.0634375\\
632	40.0925\\
};
\addplot [color=mycolor4,mark size=2.0pt,only marks,mark=o,mark options={solid}]
  table[row sep=crcr]{%
2253	14.7853125\\
2222	15.27625\\
2191	15.7478125\\
2162	16.215\\
2133	16.6640625\\
2077	17.5246875\\
1999	18.740625\\
1951	19.51\\
1927	19.8721875\\
1904	20.23\\
1860	20.925\\
1758	22.524375\\
1739	22.824375\\
1632	24.48\\
1584	25.245\\
1495	26.6296875\\
1467	27.0478125\\
1176	31.605\\
1032	33.8625\\
632	40.0925\\
};
\addlegendentry{TBM (d=5)};

\addplot [color=mycolor5,solid,line width=0.75pt,forget plot]
  table[row sep=crcr]{%
1	0.005625\\
2352	13.23\\
1	0.00625\\
2285	14.28125\\
1	0.0065625\\
2253	14.7853125\\
1	0.0071875\\
2191	15.7478125\\
1	0.0075\\
2162	16.215\\
1	0.008125\\
2105	17.103125\\
1	0.0090625\\
2025	18.3515625\\
1	0.01\\
1951	19.51\\
1	0.0103125\\
1927	19.8721875\\
1	0.0109375\\
1882	20.584375\\
1	0.0146875\\
1649	24.2196875\\
1	0.0175\\
1509	26.4075\\
1	0.0325\\
1038	33.735\\
};
\addplot [color=mycolor5,mark size=2.0pt,only marks,mark=square,mark options={solid}]
  table[row sep=crcr]{%
2352	13.23\\
2285	14.28125\\
2253	14.7853125\\
2191	15.7478125\\
2162	16.215\\
2105	17.103125\\
2025	18.3515625\\
1951	19.51\\
1927	19.8721875\\
1882	20.584375\\
1649	24.2196875\\
1509	26.4075\\
1038	33.735\\
};
\addlegendentry{TBM (d=6)};

\addplot [color=mycolor6,solid,line width=0.75pt,forget plot]
  table[row sep=crcr]{%
1	0.0053125\\
2388	12.68625\\
1	0.0059375\\
2318	13.763125\\
1	0.00625\\
2285	14.28125\\
1	0.0071875\\
2191	15.7478125\\
1	0.00875\\
2051	17.94625\\
1	0.01\\
1951	19.51\\
1	0.0171875\\
1523	26.1765625\\
};
\addplot [color=mycolor6,mark size=2.0pt,only marks,mark=diamond,mark options={solid}]
  table[row sep=crcr]{%
2388	12.68625\\
2318	13.763125\\
2285	14.28125\\
2191	15.7478125\\
2051	17.94625\\
1951	19.51\\
1523	26.1765625\\
};
\addlegendentry{TBM (d=7)};

\addplot [color=mycolor7,solid,line width=0.75pt,forget plot]
  table[row sep=crcr]{%
1	0.005\\
2424	12.12\\
1	0.0059375\\
2318	13.763125\\
1	0.0096875\\
1975	19.1328125\\
};
\addplot [color=mycolor7,mark size=2.0pt,only marks,mark=star,mark options={solid}]
  table[row sep=crcr]{%
2424	12.12\\
2318	13.763125\\
1975	19.1328125\\
};
\addlegendentry{TBM (d=8)};

\addplot [color=mycolor8,solid,line width=0.75pt,forget plot]
  table[row sep=crcr]{%
1	0.0046875\\
2461	11.5359375\\
};
\addplot [color=mycolor8,mark size=2.0pt,only marks,mark=triangle,mark options={solid}]
  table[row sep=crcr]{%
2461	11.5359375\\
};
\addlegendentry{TBM (d=9)};

\end{axis}
\end{tikzpicture}%
\caption{Achievable sum-DoF per channel use ($\dof_{\textrm{TBM}}(\Ka)/T$) vs. $\Ka$ for different tensor sizes ($d$ and $T_i$), for $T=3200$ and $N=50$. The markers denote the case $\Ka = \overline{R} -1$, while the slope of the lines going through the origin represents the per-user DoF.}
\label{fig:dof}
\end{figure}

Figure~\ref{fig:dof} depicts the sum-DoF per channel use achievable using different tensor sizes; specifically, for fixed values of $T$ and $N$, it depicts the points $(\Ka,\frac{\dof_{\textrm{TBM}}(\Ka)}{T})$ for values of $\Ka$ ranging from 1 to $\overline{R}-1$, achievable using various choices of the factorization $T = \prod_{i=1}^d T_i$.
The bound \eqref{eq:dof_upperbound} and the cooperative upper bound are also depicted.
Note that this figure highlights an interesting trade-off between spectral efficiency and the maximum degree of contention: in the regime of large $\Ka$ (i.e. when \eqref{eq:dof_upperbound} is tight), using a tensor size supporting a higher number of users results in a decrease in both sum-DoF and per-user DoF.
TBM configurations achieving the highest per-user DoF (on the left of the figure) correspond to unbalanced tensors (eqs.~\eqref{eq_unbalanced1}-\eqref{eq_unbalanced2}); in this case, the upper bound $\Ka< \overline{R}$ is more restrictive than \eqref{eq:dof_upperbound}.

\section{Simulation results}
\label{section_simul}

The performance of the proposed TBM has been evaluated through simulations.
Our simulation includes a Bose–Chaudhuri–Hocquenghem (BCH) binary code applied to the messages transmitted by each user.
In addition to its error correction capability, the binary code allows for more refined multi-user decoding strategies (such as the turbo-like approach used below).
In order to facilitate the comparison with results available in the literature, we assume here that the number of active users is known to the receiver through a genie, and therefore set $\overline{\Ka} = \Ka$ in our simulations.
Note however that error detection can be instrumental in the case where $\Ka$ is unknown a priori.

\subsection{Modulation Parameters}
\label{sec:binary_code}
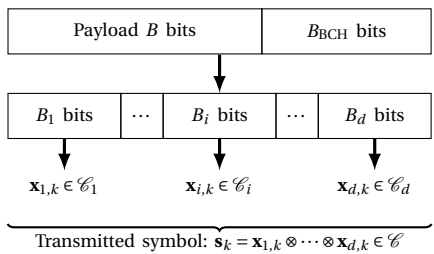
\begin{figure}[h]
\centering
\begin{tikzpicture}[scale=0.75, every node/.style={transform shape}]
\tikzstyle{quadri}=[rectangle,draw, text centered, minimum height= 0.8cm]
\tikzstyle{textcen}=[text centered, minimum height= 0.5cm]
\tikzstyle{link}=[->,>=latex,very thick]
\node[minimum height= 0.8cm, thick] (Pcenter) at (2.5,3) {};
\node[quadri, minimum width=4.5cm] (P) at (1,3) {Payload $B$ bits};
\node[quadri, minimum width=3cm] (CRC) at (4.75,3) {$\BBCH$ bits};
\node[quadri, minimum width=2cm] (P1) at (-0.25,1.5) {$B_1$ bits};
\node[quadri, minimum width=0.75cm] at (1.125,1.5) {$\dots$};
\node[quadri, minimum width=2cm] (Pi) at (2.5,1.5) {$B_i$ bits};
\node[quadri, minimum width=0.75cm] at (3.875,1.5) {$\dots$};
\node[quadri, minimum width=2cm] (Pd) at (5.25,1.5) {$B_d$ bits};
\node[textcen, minimum  width=2cm] (x1) at (-0.25,0.25) {$\x_{1,k}\in\mathcal{C}_1$};
\node[textcen, minimum width=2cm] (xi) at (2.5,0.25) {$\x_{i,k}\in\mathcal{C}_i$};
\node[textcen, minimum width=2cm] (xd) at (5.25,0.25) {$\x_{d,k}\in\mathcal{C}_d$};
\node[textcen, minimum width=2cm] (x) at (2.5,-0.75) {Transmitted symbol: $\s_k = \x_{1,k}\otimes \dots\otimes \x_{d,k}\in\mathcal{C}$};
\draw[link] (Pcenter.south)--(Pi);
\draw[link] (P1)--(x1);
\draw[link] (Pi)--(xi);
\draw[link] (Pd)--(xd);
\draw [thick, decoration={brace, mirror, raise=0.5cm}, decorate] (x1.west) -- (xd.east);
\end{tikzpicture}
\caption{Bit mapping for user $k$}
\label{fig:bit_distribution}
\end{figure}

Our set-up assumes a payload of $B=96$ information bits at each transmitter (referred to as a message), which are BCH-encoded into a codeword of length $\Btot=110$ bits; the BCH code can correct up to $2$ bit errors thanks to the $\BBCH=14$ bits of redundancy.
Bit-to-symbol mapping is performed as depicted on Fig.~\ref{fig:bit_distribution}, i.e. the $\Btot$ coded bits are split into $d$ sets of respectively $B_1,\dots,B_d$ bits, corresponding to the $d$ tensor dimensions.
The $i$-th set, comprised of $B_i$ bits, is mapped to an element of the sub-constellation $\mathcal{C}_i$; we used the Grassmannian constellation design from \cite{TWC2020_cubesplit} for the $\mathcal{C}_i$, due to the availability of a low-complexity approximate demapper.
Finally, the vector symbol $\s_k$ is formed by computing the Kronecker product \eqref{eq:kronecker}.

We consider two tensor sizes for the TBM, namely $(T_1,T_2)=(64,50)$ and $(T_1,T_2,T_3,T_4,T_5)=(8,5,5,4,4)$.
According to \cite[Lemma~1]{TWC2020_cubesplit},  the minimum distance between the elements of $\mathcal C_i$ is maximized when the $B_i$ are proportional to $T_i-1$.
In order to approximately fulfill this requirement, the $\Btot = B+\BBCH=110$ bits are split according to $(B_1,B_2)=(62,48)$ and $(B_1,B_2,B_3,B_4,B_5)=(37,21,21,16,15)$ respectively for the two considered cases.
In both cases, the dimension of $\s_k$ is $T = \prod_{i=1}^d T_i = 3200$ channel uses.

\subsection{Receiver Details}
\label{sec:receiver_details}
Taking into account the non-convexity of the objective function of \eqref{eq:cpd}, the receiver used in the simulations consists of two iterations of the two-step decoder described in Section~\ref{sec:twostep_decoder}.  %
At the first iteration, user separation \eqref{eq:cpd} and demapping \eqref{eq:chordal} are performed in order to recover the coded binary stream; then, for each binary vector fulfilling exactly the BCH constraints, the corresponding message is deemed valid.
Thus, at the first iteration, the BCH code is used for error detection only.
At iteration 2, user separation \eqref{eq:cpd} is performed a second time, during which the symbols corresponding to the messages decoded at iteration 1 are excluded from the optimization variables and replaced by their hard decision values.
After user separation and demapping, the binary vectors are BCH decoded: if decoding is successful (now using the error correction capability of the BCH code), the corresponding message is deemed valid; if decoding fails, the vector is discarded.
Let $\hat{\pazocal{L}}$ denote the list of messages deemed valid at either of the two iterations, while $\pazocal{L}$ denotes the list of messages actually transmitted by the active users.
Since the messages can be decoded only up to a permutation over the users, the considered error metric is the average message error ratio (MER),
\begin{equation}
\label{eq:perusererror}
\text{MER} = \mathbb{E}\Bigg[\min\Big(
\underbrace{\frac{|\pazocal{L}\setminus\hat{\pazocal{L}}|}{|\pazocal{L}|}}_{\substack{\text{average ratio}\\ \text{of missed messages}}} +
\underbrace{\frac{|\hat{\pazocal{L}}\setminus\pazocal{L}|}{|\hat{\pazocal{L}}|}}_{\substack{\text{average ratio}\\ \text{of phantom messages}}}
 ,1\Big)\Bigg].
\end{equation}
The MER accounts for two types of error events: (i) a transmitted message was not detected; (ii) a detected message was not transmitted.
The two error events may be correlated, therefore we limit the MER value to at most 1.

In our implementation, the approximate CPD \eqref{eq:cpd} is implemented using the nonlinear least square algorithm combined with a preconditioner proposed in \cite{sorber13}.

\subsection{Performance Results -- Unsourced Scenario}

The results in this section have been obtained for complex Gaussian i.i.d. (across the users and the receive antennas) fading channels with unit variance.
The entries of the noise $\w$ are i.i.d complex Gaussian with variance $\sigma^2$, while the transmitted symbols are normalized according to the considered energy per bit to noise ratio $E_b/N_0=\frac{\|\s_k\|^2}{B\sigma^2}$ for all $k$.

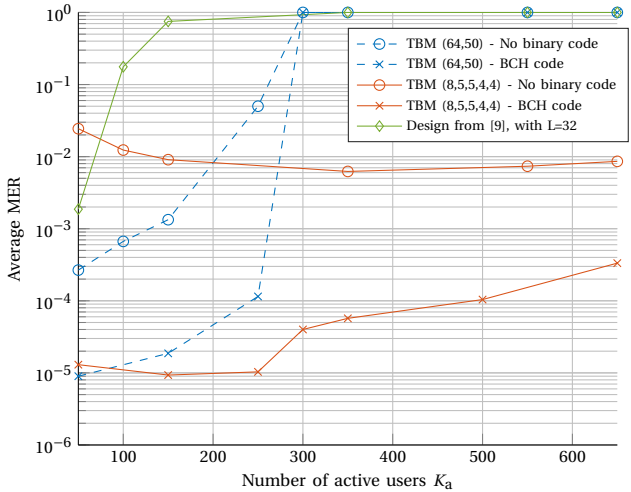
\begin{figure}[h!]
\centering
\scriptsize
%
%
\definecolor{mycolor1}{rgb}{0.00000,0.44700,0.74100}%
\definecolor{mycolor2}{rgb}{0.85000,0.32500,0.09800}%
\definecolor{mycolor3}{rgb}{0.92900,0.69400,0.12500}%
\definecolor{mycolor4}{rgb}{0.49400,0.18400,0.55600}%
\definecolor{mycolor5}{rgb}{0.46600,0.67400,0.18800}%
\begin{tikzpicture}

\begin{axis}[%
width=2.820833in,
height=2.265625in,
at={(0.758333in,0.48125in)},
scale only axis,
xmin=50,
xmax=650,
xlabel={Number of active users $\Ka$},
xmajorgrids,
ymode=log,
ymin=1e-06,
ymax=1,
ylabel={Average MER},
ymajorgrids,
yminorgrids,
title style={font=\bfseries},
axis x line*=bottom,
axis y line*=left,
legend style={at={(0.5,0.7)},anchor=south west,legend cell align=left,align=left,draw=white!15!black,nodes={scale=0.8, transform shape}}
]

\addplot [color=mycolor1,dashed,mark=o,mark options={solid}]
  table[row sep=crcr]{%
50	0.000266666666666667\\
100	0.000666666666666667\\
150	0.00133333333333333\\
250	0.05000000000000000\\
300	1\\
350	1\\
550	1\\
650	1\\
};
\addlegendentry{TBM (64,50) - No binary code};




\addplot [color=mycolor1,dashed,mark=x,mark options={solid,rotate=180}]
  table[row sep=crcr]{%
50	9.00000000000000e-06\\
150	1.86666666666666e-05\\
250 1.14880000000000e-04\\
300	1\\
550	1\\
650	1\\
};
\addlegendentry{TBM (64,50) - BCH code};

\addplot [color=mycolor2,solid,mark=o,mark options={solid}]
  table[row sep=crcr]{%
50	0.0244\\
100	0.0123333333333333\\
150	0.00906666666666667\\
350	0.00622857142857143\\
550	0.00735757575757576\\
650	0.00859487179487179\\
};
\addlegendentry{TBM (8,5,5,4,4) - No binary code};




\addplot [color=mycolor2,solid,mark=x,mark options={solid,rotate=180}]
  table[row sep=crcr]{%
50	1.30000000000000e-05\\
150	0.93333333333333e-05\\
250	1.03333333333333e-05\\
300	4.00000000000000e-05\\
350	5.711400000000000e-05\\
500	1.040200000000000e-04\\
650	3.330200000000000e-04\\
};
\addlegendentry{TBM (8,5,5,4,4) - BCH code};

\addplot [color=mycolor5,solid,mark=diamond,mark options={solid}]
  table[row sep=crcr]{%
50	0.00186666666666667\\
100	0.1762\\
150	0.750755555555556\\
350	1\\
550	1\\
650	1\\
};
\addlegendentry{Design from \cite{fengler19}, with L=32};



\end{axis}
\end{tikzpicture}%
\caption{Average per-user MER vs. number of active users for a BS with $N=50$ antennas and $E_b/N_0=0$ dB.}
\label{fig:perf1}
\end{figure}

In Figure~\ref{fig:perf1}, the performance of TBM is compared to the compressed sensing--based approach proposed in \cite{fengler19}, for $\Ka$ ranging from $50$ to $650$, and $E_b/N_0 = 0$ dB.
It can be observed that tensor-based methods maintain a low MER for $\Ka$ ranging up to 650, while the design from \cite{fengler19} exhibits a large MER already for $\Ka\approx 100$.
In order to illustrate the role of the binary code, we also consider an \emph{uncoded} set-up, not involving any binary code ($\BBCH=0$).
In this case, a single decoding stage (eqs.~\eqref{eq:cpd}-\eqref{eq:chordal}) is performed at the receiver, and $\hat{\pazocal{L}}$ always consists of $\Ka$ messages).
Also observe that TBM with sizes $(8,5,5,4,4)$ yields consistently superior performance to what is achieved using tensors of sizes $(64,50)$, which seems to indicate that, for a fixed $T$, designs with higher $d$ are preferable.

\begin{figure}[h!]
\centering
\scriptsize
%
%
\definecolor{mycolor1}{rgb}{0.00000,0.44700,0.74100}%
\definecolor{mycolor2}{rgb}{0.85000,0.32500,0.09800}%
\definecolor{mycolor3}{rgb}{0.92900,0.69400,0.12500}%
\definecolor{mycolor4}{rgb}{0.49400,0.18400,0.55600}%
\definecolor{mycolor5}{rgb}{0.46600,0.67400,0.18800}%
\begin{tikzpicture}

\begin{axis}[%
width=2.520833in,
height=1.565625in,
at={(0.758333in,0.48125in)},
scale only axis,
unbounded coords=jump,
xmin=10,
xmax=700,
xlabel={Number of active users},
xmajorgrids,
xminorgrids,
ymin=-13,
ymax=25,
ylabel={$E_b/N_0$ (dB)},
ymajorgrids,
yminorgrids,
title style={font=\bfseries},
axis x line*=bottom,
axis y line*=left,
legend style={at={(0.4,0.23)},anchor=south west,legend cell align=left,align=left,draw=white!15!black,nodes={scale=0.8, transform shape}}
]

\addplot [color=mycolor5,solid,mark=diamond,mark options={solid}]
  table[row sep=crcr]{%
10	-3\\
25	-2\\
50	-1\\
100	1\\
150	9\\
200	100\\ 
};
\addlegendentry{Design from \cite{fengler19} with 32 subblocks, $N=50$, Rayleigh};

\addplot [color=mycolor2,solid,mark=x,mark options={solid,rotate=180}]
  table[row sep=crcr]{%
10	-6.5\\
25	-6.5\\
50	-6.5\\
100	-7\\
200	-7\\
300	-6.5\\
500	-6\\
650	-6\\
700	100\\   
};
\addlegendentry{TBM (8,5,5,4,4) + BCH code, $N=50$, Rayleigh ($\overline{R}=2254$)};

\addplot [color=mycolor1,mark=x,mark options={solid,rotate=180}]
  table[row sep=crcr]{%
10	-10.5\\
50	-10.5\\
100	-9.5\\
150	-8.5\\
200	-6\\
225	-5\\
250	-3\\
275	0\\
300	100\\ 
};
\addlegendentry{TBM (64,50) + BCH code, $N=50$, Rayleigh ($\overline{R}=988$)};

\addplot [color=mycolor2,solid,mark=x,dashed,mark options={solid,rotate=180}]
  table[row sep=crcr]{%
10	17\\
25	17\\
50	18\\
60	18\\
80	18\\
85	100\\ 
};
\addlegendentry{TBM (8,5,5,4,4) + BCH code, $N=1$, Rayleigh  ($\overline{R}=146$)};

\addplot [color=black,dashed,mark options={solid}]
  table[row sep=crcr]{%
10	8.1162\\
25	8.1162\\
50	8.1162\\
100	8.3166\\
200	8.4168\\
300 14.7295\\
500	28.4569\\
700	42.9860\\
};
\addlegendentry{Fano-type Converse, $N=1$, Rayleigh};

\addplot [color=mycolor4,dashed,mark options={solid,rotate=180}]
  table[row sep=crcr]{%
10	12.3848\\
25  12.5761\\
50  12.8867\\
100 13.4904\\
200 15.7532\\
300 23.0762\\
400 33.4370\\
500 38.2363\\
650 64.4052\\
};
\addlegendentry{Random coding achievability, $N=1$, Rayleigh};

\addplot [color=mycolor2,dashed,mark=triangle,mark options={solid,rotate=180}]
  table[row sep=crcr]{%
10	9\\
25	10\\
50	10\\
60	11\\
75	100\\ 
};
\addlegendentry{TBM (8,5,5,4,4) + BCH code, $N=1$, AWGN ($\overline{R}=146$)};


  
\addplot [color=mycolor4,dashed,mark=triangle,mark options={solid,rotate=180}]
  table[row sep=crcr]{%
10	2.0408\\
25  2.0408\\
50  4.0816\\
100 10.2041\\
200 24.4898\\
650 93.8776\\
};
\addlegendentry{Random coding achievability, $N=1$, AWGN channel};

\end{axis}
\end{tikzpicture}%
\caption{Minimum $E_b/N_0$ required to achieve $\text{PUPE}\leq 0.1$ vs. number of active users for $T=3200$.}
\label{fig:perf2}
\end{figure}
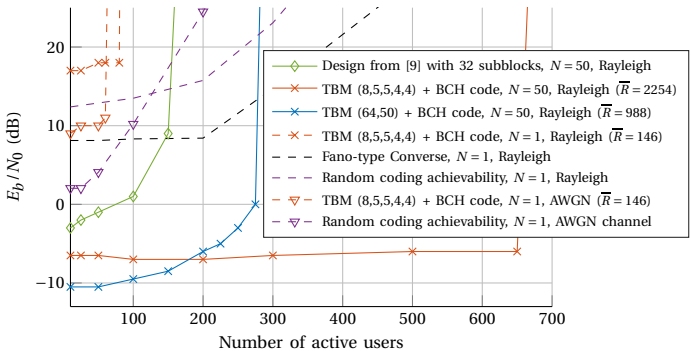

Figure~\ref{fig:perf2} depicts the $E_b/N_0$ required to achieve a target error rate, for $N=50$ and $N=1$ antennas.
To facilitate comparison with the existing results, we now focus on the average per user probability of error (PUPE) \cite{polyanskyi17}, 
defined as
\begin{equation}
\text{PUPE} = \mathbb{E}\Bigg[\frac{|\pazocal{L}\setminus\hat{\pazocal{L}}|}{|\pazocal{L}|}\Bigg],
\end{equation}
and depict the $E_b/N_0$ required to achieve a PUPE lower or equal to $0.1$ 
For TBM, the value of $\overline{R}$ is included in the legend.
We compare to the approach from \cite{fengler19} with 32 subblocks.
We also include the single-antenna receiver case for comparison, for which theoretical bounds are available, such as a Fano-type converse bound and a random coding achievability bound, both from \cite[Appendix B]{kowshikpolyanskyi19b} in the Rayleigh fading case, and the achievability bound of \cite{polyanskyi17} for the AWGN case.
We observe that TBM benefits from spatial diversity (the number of users that can be successfully decoded increases with $N$, as suggested by the DoF analysis), and allows to achieve very high multiplexing gains for the case of multiple receive antennas, with reasonable packet loss rate and $E_b/N_0$.
For instance, for $N=50$, it allows up to 650 active users over a Rayleigh fading channel, which translates into a spectral efficiency of $19.5$ bits/channel access.

\subsection{Performance Results -- Sourced Scenario}

\begin{figure}[h!]
\centering
\scriptsize
%
%
\definecolor{mycolor1}{rgb}{0.00000,0.44700,0.74100}%
\definecolor{mycolor2}{rgb}{0.85000,0.32500,0.09800}%
\definecolor{mycolor3}{rgb}{0.92900,0.69400,0.12500}%
\definecolor{mycolor4}{rgb}{0.49400,0.18400,0.55600}%
\definecolor{mycolor5}{rgb}{0.46600,0.67400,0.18800}%
\begin{tikzpicture}

\begin{axis}[%
width=2.820833in,
height=2.265625in,
at={(0.758333in,0.48125in)},
scale only axis,
xmin=-12,
xmax=0,
xlabel={$E_b/N_0$ (dB)},
xmajorgrids,
ymode=log,
ymin=1e-06,
ymax=1,
ylabel={Average PER},
ymajorgrids,
yminorgrids,
title style={font=\bfseries},
axis x line*=bottom,
axis y line*=left,
legend style={at={(0.01,0.18)},anchor=south west,legend cell align=left,align=left,draw=white!15!black,nodes={scale=0.8, transform shape}}
]

\addplot [color=mycolor1,dashed,mark=o,mark options={solid}]
  table[row sep=crcr]{%
-12	 0.9920\\
-11	0.8870\\
-10	0.2950\\
-9	0.0600\\
-8	0.0134\\
-6	0.0034\\
-4  0.0012\\
-2  7.00e-4\\ 
0   2.90e-4\\ 
};
\addlegendentry{TBM (64,50) - No binary code};

\addplot [color=mycolor1,dashed,mark=x,mark options={solid,rotate=180}]
  table[row sep=crcr]{%
-12	0.9620\\
-11	0.4280\\
-10	0.0390\\
-9	0.0050\\
-6  0.7500e-5\\ 
-4  0.500e-5\\ 
-2  0.500e-5\\
0   0.500e-5\\
};
\addlegendentry{TBM (64,50) - BCH code};

\addplot [color=mycolor2,solid,mark=o,mark options={solid}]
  table[row sep=crcr]{%
-12	 0.9710\\
-11	0.9320\\
-10	0.8440\\
-9	0.7030\\
-8	0.5940\\ 
-6	0.2350\\
-4  0.0690\\
-2  0.0210\\
0   0.0070\\
};
\addlegendentry{TBM (8,5,5,4,4) - No binary code};

\addplot [color=mycolor2,solid,mark=x,mark options={solid,rotate=180}]
  table[row sep=crcr]{%
-12	0.9490\\
-11	0.8690\\
-10	0.6870\\
-9	0.4660\\
-8	0.2810\\
-6	0.0750\\
-4	0.0198\\
-2	0.0028\\
0	2.40e-4\\
};
\addlegendentry{TBM (8,5,5,4,4) - BCH code};

\addplot [color=mycolor5,solid,mark=diamond,mark options={solid}]
  table[row sep=crcr]{%
-12	 1\\
-11	 1\\
-10	 1\\
-9	 1\\
-8	 1\\
-6   0.9970\\
-4   0.8990\\
-2   0.5330\\
0	 0.1290\\
};
\addlegendentry{Design from \cite{fengler19}, with L=32};

\end{axis}
\end{tikzpicture}%
\caption{PER vs. $E_b/N_0$ for $T=3200$, $N=50$, $\Ka=100$ and $K=8192$.}
\label{fig:perf3}
\end{figure}
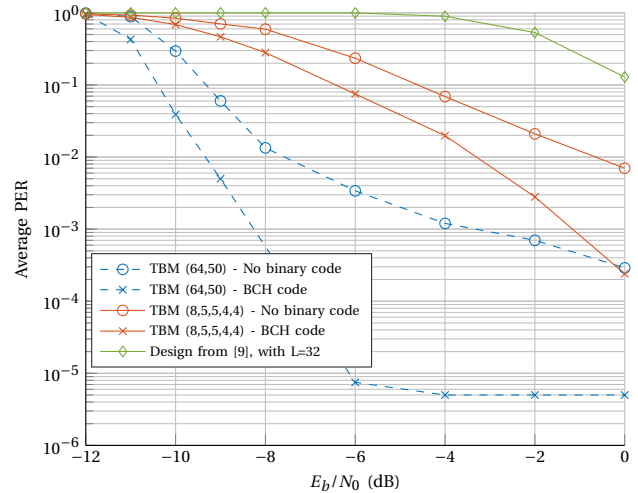

Let us now consider the case where every transmitter embeds its identity (ID) in the payload, in addition to the $B=96$ information bits.
We assume that $\Ka=100$ users are active, out of a total of $K=8192$ users; therefore, $\BID=\log_2(K)=13$ bits are required to encode the user ID.
Hence, each user transmits a total of $\Btot=\BID+B=109$ bits when no binary code is used and $\Btot=\BID+B+\BBCH=123$ bits when the BCH code mentioned in Section~\ref{sec:binary_code} is used. 
At the receiver side, in addition to the unsourced decoding procedure described in Section~\ref{sec:receiver_details}, the transmitting users are identified based on the ID embedded in the decoded payload.
This allows to evaluate the packet error rate (PER) metric as in a sourced scenario.
If several messages with the same user ID are decoded, they are discarded and counted as an error.
The resulting PER (averaged over the active users) is depicted in Figure~\ref{fig:perf3}; the error floor at high $E_b/N_0$ is due to the lack of optimality of the solution to the non-convex problem \eqref{eq:cpd}.
It is noticeable that, when using BCH code and in the case of $\Ka=100$ active users, TBM $(8,5,5,4,4)$ PER is higher than TBM $(64,50)$ in the sourced scenario while they have similar MER in the unsourced scenario, hence the former is more impacted by the addition of the $\BID$ identity bits.
This may be explained by the number of DoF which is higher in the latter making its use more suitable for denser constellations.

\bibliographystyle{IEEEtran}
\bibliography{refs}

\newpage
\appendices
\section{Details on the Single-User Decoder}
\label{app:single-user}
Fixing $\x_{i,k}\in\mathcal{C}_i$ for all $i,k$, and solving \eqref{eq:single-user} with respect to $\h_k$,  yields
\begin{equation}
\label{eq:hk}
\h_k^* = \frac{(\x_{1,k}\otimes\dots\otimes \x_{d,k})^H (\hat{\z}_{1,k}\otimes\dots\otimes \hat{\z}_{d,k}) }{\|\x_{1,k}\otimes\dots\otimes \x_{d,k}\|^2}\hat{\h}_k.
\end{equation}
Substituting \eqref{eq:hk} into the objective function of \eqref{eq:single-user}, and applying the property that $\a\otimes \b=\text{vec}\big(\a\b^T\big)$ for arbitrary vectors $\a$ and $\b$ to both terms, we get
\begin{eqnarray}
\nonumber&&\| \hat{\z}_{1,k}\otimes\dots\otimes \hat{\z}_{d,k}\otimes\hat{\h}_k - \x_{1,k}\otimes\dots\otimes \x_{d,k}\otimes\h_k^*\|\\
\nonumber&=&\big\| \text{vec}\big((\hat{\z}_{1,k}\otimes\dots\otimes \hat{\z}_{d,k})\hat{\h}_k^T\big) - \text{vec}\big((\x_{1,k}\otimes\dots\otimes \x_{d,k})(\h_k^*)^T\big)\big\|\\
&=&\left\|P_{\x} (\hat{\z}_{1,k}\otimes\dots\otimes \hat{\z}_{d,k}) \hat{\h}_k^T\right\|_2^2\label{eq:projection},
\end{eqnarray}
where we define the projection matrix $P_{\x}=\I_T-\frac{(\x_{1,k}\otimes\dots\otimes \x_{d,k})(\x_{1,k}\otimes\dots\otimes \x_{d,k})^H}{\|\x_{1,k}\otimes\dots\otimes \x_{d,k}\|^2}$.
Using $\|\a\b^T\|_2^2=\|\a\|^2\|\b\|^2$ and the fact that $P_{\x}$ is a projector, the optimization problem \eqref{eq:single-user} is equivalent to
\begin{equation}
\max_{\substack{\x_{i,k}\in\mathcal{C}_i, \ \forall i}} \frac{\left\|(\x_{1,k}\otimes\dots\otimes \x_{d,k})(\x_{1,k}\otimes\dots\otimes \x_{d,k})^H (\hat{\z}_{1,k}\otimes\dots\otimes \hat{\z}_{d,k})\right\|_2^2}{\|\x_{1,k}\otimes\dots\otimes \x_{d,k}\|^4}.\\
\end{equation}
Finally, using $(\a\otimes\b)^H(\a'\otimes\b')= (\a^H\a')(\b^H\b')$ we obtain
\begin{equation}
\max_{\x_{i,k}\in\mathcal{C}_i, \ \forall i} \prod_{i=1}^d \frac{\left|\x_{i,k}^H\hat{\z}_{i,k}\right|}{\|\hat{\z}_{i,k}\| \|\x_{i,k}\|}
\end{equation}
which is clearly separable, and yields \eqref{eq:chordal}.

\end{document}